\documentclass{emulateapj}

\newcommand\x         {\hbox{$\times$}}

\newcommand{\Lisol}{\mbox{$L_{I_c, \odot}$}}
\newcommand{\MIsol}{\mbox{$M_{I_c,\odot}$}}

\def\u                {{\em u}}
\def\g                {{\em g}}
\def\r                {{\em r}}
\def\gmag                {{\em g}}
\def\rmag                {{\em r}}
\def\imag                {{\em i}}
\def\z                {{\em z}}

\def\V                {{\em V}}
\def\I                {$I_c$}
\def\dph              {DoPHOT}
\def\ph               {Photo}
\def\Mo               {\hbox{$M_{\odot}$}}

\def\sqdegs           {\hbox{$\Box^\circ$}}
\def\sqdeg           {\hbox{$\Box^\circ$}}
\def\deg              {\hbox{$^\circ$}}

\def\leo              {Leo~I}
\def\dSph             {dSph}
\def\etal             {et al.}
\def\kms              {km~s$^{-1}$}
\def\comm#1  { {\tt (COMMENT: #1)} }

\slugcomment{Version of 16 June 2006}

\shorttitle{SDSS crowded field photometry: \leo}
\shortauthors{V. Smol\v{c}i\'{c} \etal}

\begin{document}

\title{Improved photometry of SDSS crowded field images: Structure and dark
  matter content in the dwarf spheroidal galaxy Leo~I} 

\author{V.~Smol\v{c}i\'{c}}
\affil{Max-Planck-Institut f\"ur Astronomie, K\"onigstuhl 17, Heidelberg,
  D-69117, Germany} 

\author{D.~Zucker}
\affil{Institute of Astronomy, University of Cambridge, Madingley Road,
  Cambridge CB3 0HA, UK} 

\author{E.~F.~Bell, M.~G.~Coleman, H.~W.~Rix, E.~Schinnerer}
\affil{Max-Planck-Institut f\"ur Astronomie, K\"onigstuhl 17, Heidelberg,
  D-69117, Germany} 

\author{\v{Z}.~Ivezi\'{c}}
\affil{Department of Astronomy, University of Washington, Box 351580, Seattle,
  WA 98195-1580, USA }

\and

\author{A.~Kniazev}
\affil{South African Astronomical Observatory, Observatory Road, Cape Town,
  South Africa}

\begin{abstract}

  We explore how well crowded field point-source photometry can be
  accomplished with Sloan Digital Sky Survey (SDSS) data. For this purpose, we
  present a photometric pipeline based on \dph\ \citep{schechter93}, and
  tuned for analyzing crowded-field images from the SDSS. Using Monte Carlo
  simulations we show that the completeness of source extraction is above
  $80\%$ to an $i$ band AB magnitude of $\lesssim21$ and a stellar surface
  density of $\sim200$~arcmin$^{-2}$.  Hence, a specialized data pipeline can
  be efficiently used for fairly crowded fields, such as nearby resolved
  galaxies in SDSS images, where the standard SDSS photometric package \ph,
  when applied in normal survey mode, gives poor results.
  
  We apply our pipeline to an area of $\sim3.55$\sqdeg\ around the dwarf
  spheroidal galaxy (\dSph) \leo. Using the resulting multi-band (\g,\r,{\em
    i}) photometry we construct a high signal-to-noise star-count map of \leo,
  utilizing an optimized filter in color-magnitude space. This filter reduces
  the foreground contamination by $\sim80\%$ and enhances the central stellar
  surface density contrast of the dwarf by a factor of $\gtrsim4$, making this
  study the deepest wide-field study of the \leo\ \dSph\ based on accurate CCD
  photometry. We find that the projected spatial structure of \leo\ is
  ellipsoidal. The best fitting empirical King model to the stellar-surface
  density profile yields core and tidal radii of $(6.21\pm0.95)'$ and
  $(11.70\pm0.87)'$, respectively. This corresponds to $(460\pm75)$~pc and
  $(860\pm86)$~pc assuming a distance to \leo\ of $254^{+19}_{-16}$~kpc.
  The radial surface-density profile deviates from the King profile towards
  outer radii, yet we find no evidence for 'S' shaped or irregular tidal
  debris out to a stellar surface-density of $4\x10^{-3}$ of the central
  value. From the luminosity function of all possible \leo\ stars, which we
  carefully extrapolated to faintest magnitudes, we determine the total
  \I-band luminosity of \leo\ to be $(3.0\pm0.3)\x10^{6}\,\Lisol$. We model
  the mass of the \dSph\ using the spherical and isotropic Jeans
  equation and infer a central mass density of $0.07\,\Mo\,\mathrm{pc}^{-3}$
  leading to a central mass-to-light ratio of $\sim3$ in \I\ band solar units.
  Assuming that the mass in the system follows the distribution of the visible
  component, we constrain a lower limit on the total mass of the \dSph\ to be
  $(1.7\pm0.2)\x10^7\,\Mo$. On the other hand, if the mass in \leo\ is
  dominated by a dark-matter (DM) halo with constant density, then the mass
  within the central $12'$ yields $(2\pm0.6)\x10^8\, \Mo$. Combining the
  inferred mass estimates with the total luminosity leads to a mass-to-light
  ratio of $\gg6$ in \I\ band solar units, and possibly $>75$ if the DM halo
  dominates the mass and extends further out than $12'$. In summary, our
  results show that \leo\ is a symmetric, relaxed and bound system; this
  supports the idea that \leo\ is a dark-matter dominated system.

\end{abstract}


\section{Introduction}
\label{sec:intro}

\subsection { The Local Group as seen by the Sloan Digital Sky Survey }
Studies of the Local Group have been recently invigorated by sensitive
multi-wavelength large-area surveys, in particular the Sloan Digital Sky
Survey (SDSS; \citealt{york00,stoughton02,abaz03,abaz04,abaz05,adelmc06}). To
date SDSS has publicly released high-quality near UV to near IR five-band
photometry and accurate astrometry \citep{pier03} for $\sim$$215$ million
  objects selected over $8000$~\sqdegs (DR5; \citealt{adelmc07}).  The survey
utilizes highly-automated reduction packages. In particular, the pipeline
which extracts the multi-band photometry from SDSS images is called \ph\ (a
detailed description of the pipeline is given in \citealt{lupton02}; see also
\citealt{stoughton02}).

\begin{figure}[h!] 
\centering
\includegraphics[bb=14 14 750 603, width=\columnwidth]
                {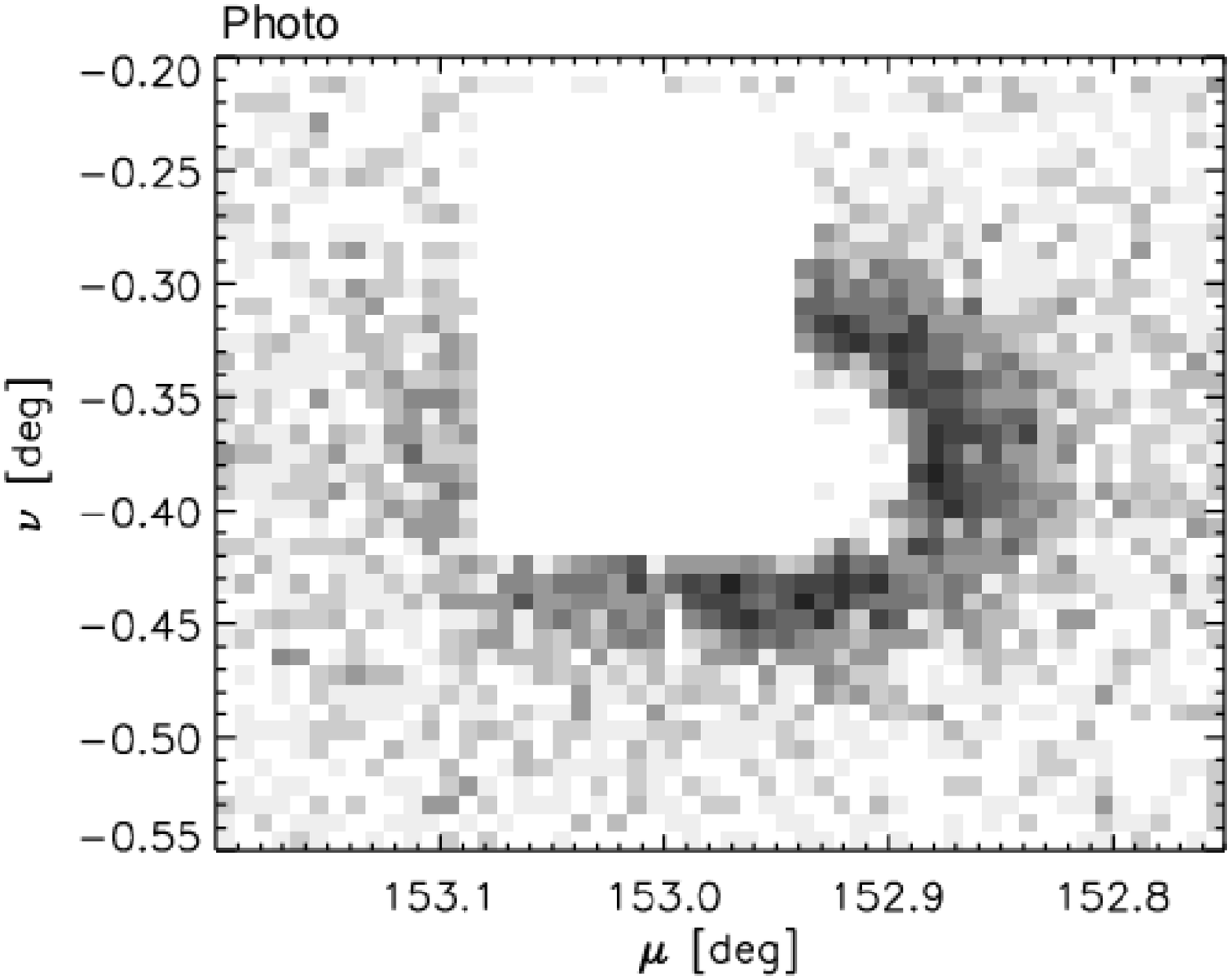}\\ 
\includegraphics[bb=14 14 750 603,, width=\columnwidth]{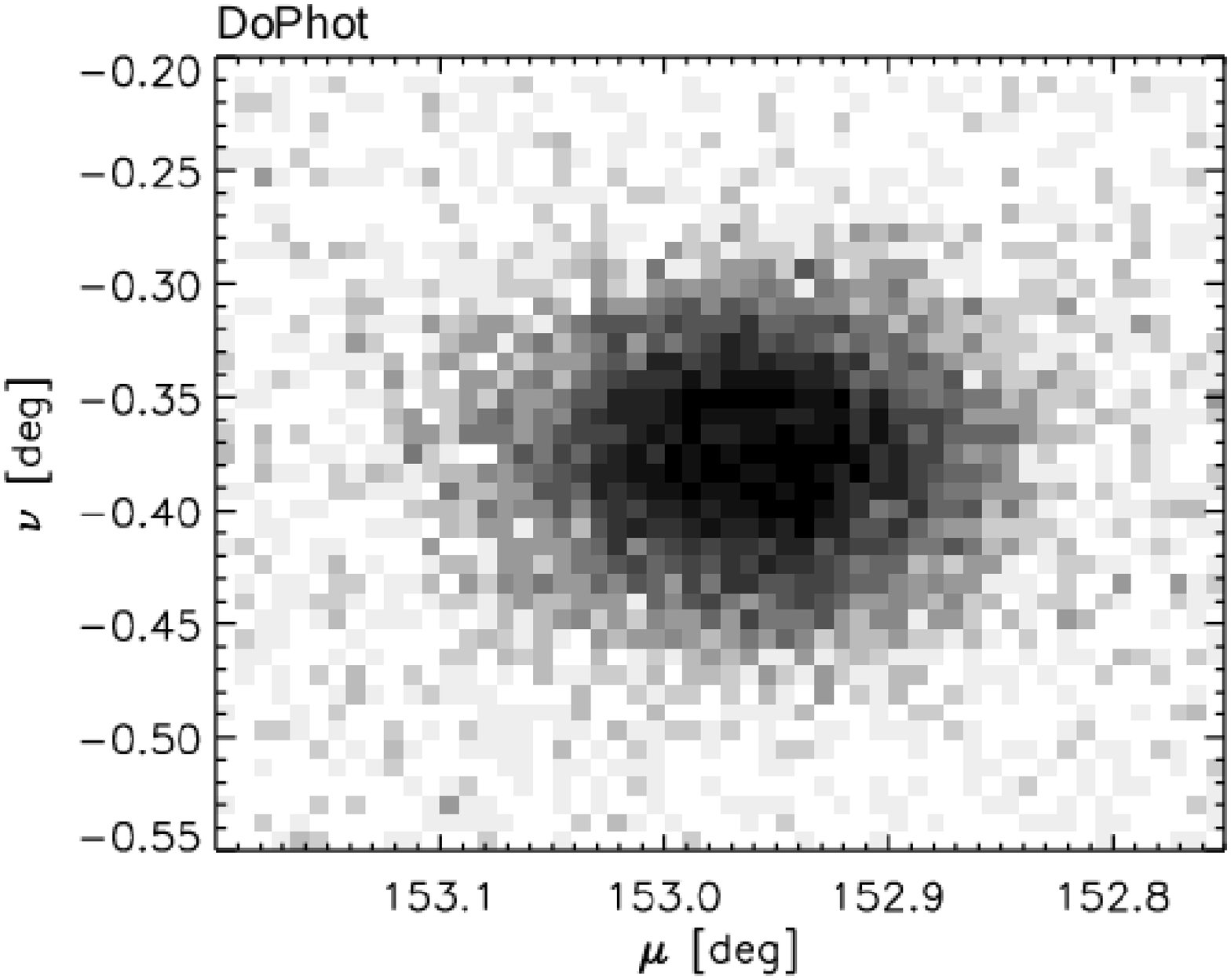}
\caption{ Stellar density plots of \leo\ shown in Great circle
    coordinates.  The grey-scale is logarithmic, and the same in both panels;
    darker colors display higher densities.  The top panel shows the SDSS
    (\ph) photometry of \leo.  The bottom panel shows \dph's photometry for
    the same area. Both panels show only objects classified as stars by each
    individual package. Note the missing stars in and around the center of the
    galaxy in the top (SDSS -- \ph) panel. }
\label{fig:leosdss}
\end{figure}

The SDSS has had a significant impact on Local Group studies.  It has led to
the discovery of the faintest known dSphs
\citep{willman05,zucker06,belokurov06a,belokurov06b,belokurov06d} as well as
to new insights into already known \dSph\ galaxies (e.g.\ Draco \dSph;
\citealt{oden01}). Analyses of the SDSS photometric dataset have revealed
tidal streams surrounding the Milky Way (for example,
\citealt{oden01,yanny03,grillmair06,belokurov06a,belokurov06b,belokurov06d}).
There have been detailed structural studies of the Galactic halo
\citep{helmi03a,xu06} and disk \citep{helmi03b,juric05} based on SDSS.
However, there is a serious problem associated with the SDSS standard
photometric pipeline: \ph\ does not extract photometry in crowded fields such
as globular clusters or nearby galaxies (see \S \ref{sec:datared} for
details). For example, the photometry is incomplete for the SDSS
fields\footnote{ An SDSS field is defined as an imaged area in the sky
  consisting of five (\u,\g,\r,{\em i},\z) frames.  } at the centre of the
\leo\ dSph galaxy (see Fig.~\ref{fig:leosdss}).


The aim of this paper is twofold.  First, we present a photometric pipeline
targeting SDSS images of crowded stellar fields.  This pipeline is based on
\dph, a software package developed by \citet{schechter93}, and is designed to
be highly automated. Secondly, we apply this pipeline to a $3.55\sqdegs$ area
centered on the \leo\ dSph, and study the properties of the dwarf.

\subsection{ \leo\ dwarf spheroidal galaxy  }
\label{sec:leointro}

The dwarf spheroidal (\dSph) galaxy \leo\ was discovered in the Palomar Sky
Survey by \citet{harrington50}.  Subsequent studies using photographic plates
were substantially hampered by the nearby ($\sim$$20'$ south of the
dSph's center) first magnitude foreground star Regulus ($\alpha$~Leonis).  As
a result, the first \leo\ color-magnitude diagram (CMD) was published only
recently \citep{fox87}.  Nevertheless, Hodge~\& Wright (1978) had already
reported the presence of an intermediate-age stellar population in the galaxy
indicated by an unusually large number of anomalous Cepheids.  Later studies
based on CCD photometry indicated that the intermediate age stars are the
dominant population \citep{rm91,lee93,demers94}.  Based on Hubble Space
Telescope (HST) observations, \citet{caputo99} and
\citet{gallart99a,gallart99b} showed that \leo\ has an extended star formation
history.  The oldest stars in \leo\ formed approximately 9 -- 13~Gyr ago, while
the youngest stars are less than 1 Gyr old.  It is this latter trait which
makes \leo\ unusual compared to most other dSphs.  An old stellar population
with an age $\gtrsim10$~Gyr has been found just recently in the outer regions
of \leo\ \citep{held00,held01}.

\leo\ is thought to be one of the most distant satellites of the Milky
Way in the Local Group. The most recent distance estimate \citep{bellazzini04}
puts the galaxy at $254^{+16}_{-19}$~kpc away from our Galaxy [they infer a
distance modulus of $(m-M)_0=22.02\pm0.13$].

Generally, the high velocity dispersions for Local Group \dSph s, combined
with dynamical mass estimates, indicate that these systems are dominated by
dark matter (DM; for a review see \citealt{m98}). The standard method for
estimating mass-to-light (M/L) ratios of pressure-supported systems is to use
one-component isotropic King models (King 1962, 1966; Richstone \& Tremaine
1986) with simplifying assumptions that the stellar velocity dispersion is
isotropic and, critically, that mass follows light.  Recent kinematic studies
have weakened this latter assumption; velocity dispersion profiles do not
demonstrate the characteristic decrease with radius predicted by isotropic
King models.  Hence it is likely that `traditional' M/L estimates for dSphs
are in fact {\em lower than the true value}.  Specifically, \citet{mateo98}
estimate the \V\ band M/L ratio of \leo\ to lie in the range of 3.5 -- 5.6 (in
solar units; see also \citealt{m98}).  This result is based on the central
velocity dispersion measured for 33 \leo\ stars using the structural
parameters (such as the core and tidal radii) given in Irwin~\& Hatzidimitriou
(1995; hereafter IH95).

IH95 determined the morphology of eight Local Group dSphs from star counts
using digitized {\em photographic plates}. One major advantage of this study
(compared to earlier studies which were largely based on eyeball counts) was
that an objective star count method was applied to all dwarfs.  However, the
usage of photographic plates is for obvious reasons much more restricted then
the usage of CCD photometry, which we utilize here.  One general problem,
which was not taken into account by IH95, is the effect of star blending,
which may be significant in regions of high number densities such as the cores
of compact stellar systems. Thus, blending may present a serious problem in
the determination of structural properties of dSphs as the number densities
are not correctly estimated.  Here we carefully approach this problem
in order to obtain the correct morphological parameters of the \leo\ dSph.  In
summary, better constraints of the structural properties of \dSph\ galaxies
are essential for more robust derivations of masses and M/L ratios.  In
particular, given the large distance from the Milky Way and the great systemic
velocity \citep{mateo98}, \leo\ plays a crucial role in the estimates of the
mass of our Galaxy (e.g.\ \citealt{zaritsky89,zaritsky99}) and the whole Local
Group \citep{lyndenbell99}.

With the advent of multi-wavelength large-area sky surveys, such as the SDSS,
the properties of nearby galaxies can be comprehensively investigated
(e.g.\ \citealt{oden01}).  In this paper we focus on deriving structural
properties of \leo\ utilizing SDSS images. We use these to constrain more
robustly the mass and total luminosity of \leo\ and finally model the mass of
the \dSph. We also address the total mass estimates for \leo\ with and without
the mass-follows-light assumption. Finally, we derive a range of possible M/L
ratios of the galaxy utilizing the improved structural parameters.

Our photometric pipeline is described in \S \ref{sec:datared}. The
extensive tests performed on the pipeline in 
order to infer the photometric accuracy and completeness are presented in
\S \ref{sec:tests}. The results on \leo\ are
given in \S\S \ref{sec:leophotom}--\ref{sec:ml}: \S\ref{sec:leophotom}
describes the color-magnitude selection of \leo\ candidates, i.e.\ the
construction of a high-contrast map of \leo, and briefly the color-magnitude
diagram (CMD) of the \dSph. We constrain the size and structure of \leo\ in \S
\ref{sec:structure} and derive the total luminosity, mass and M/L ratio in \S
\ref{sec:ml}. We discuss our results in \S \ref{sec:discussion} and summarize
them in \S \ref{sec:summary}.

\section{ The photometric pipeline  }
\label{sec:datared}

The standard SDSS photometric pipeline, \ph, has two main restrictions: (i) it
is time limited (to $1$~ms/object), and (ii) the number of extracted objects
per image cannot exceed a given number.  The photometry package therefore does
not analyze centers (and sometimes whole frames) of crowded regions (for
example, nearby resolved galaxies).  This is illustrated in Fig.
\ref{fig:leosdss}.  Here we present a pipeline developed for obtaining high
quality photometry from SDSS images of crowded fields.  The pipeline is based
on the \dph\ source extraction package \citep{schechter93} designed to search
for objects on a digital image of the sky and to extract positions, magnitudes
and classifications for those objects.  The package is widely used in the
  astronomical community and has a proven accuracy (e.g.\ 
\citealt{rm91,vogt95,gallart99b,bellazzini04}). We use a version of \dph\ 2.0
which was slightly modified by Eugene Magnier.\footnote{ The entire program is
  wrapped inside a C program which implements dynamic memory allocation,
  furthermore the C code interprets command-line arguments and allows
  compilation under f2c.}

In the following work we focus on \g, \r\ and {\em i} bands 
since the \u\ and \z\ bands are less sensitive. 
Our pipeline can be divided into four parts: 
   1) aligning the SDSS frames and extracting coefficients (from the SDSS
   tsField\footnote{ An SDSS tsField  file is a binary FITS table which
     contains parameters relevant for the  entire field (for details see
     www.sdss.org).} file) needed for further calculations, 
   2) adjusting \dph's input parameters, 
   3) running \dph\ on the individual frames and 
   4) converting the output photometry to the AB magnitude scale and computing 
      several other quantities, as well as extracting the proper astrometry
      for a given field (see \S \ref{dph_analysis} for details).
We briefly describe the pipeline below.

\subsection{ \dph's input parameters }

\dph\ requires an initial list of parameters (fine-tuned for the given image)
to run. In order to make the object-extraction from SDSS images fast and
highly automated we developed template \dph\ parameter files for each
photometric band. We fine-tuned these parameters and finally adopted the ones
giving the best detection rates.

Two sets of parameters required close attention: the `aperture box' values,
and those defining the background sky model.  Since \dph\ fits a model point
spread function (PSF) to the data (as opposed to a numerically given empirical
PSF), it also computes aperture fluxes to correct for the systematic errors
introduced by using a model PSF.  These 'aperture box' parameters were
fine-tuned to reproduce the existing SDSS photometry in un-crowded fields,
where \ph\ performs well\footnote{The parameters were tested on fields
  surrounding the \leo\ \dSph. Nonetheless, if the need for different aperture
  corrections (or any of the fixed parameters) arises, the parameters can
  easily be changed and added to the pipeline without adversely affecting its
  performance.}.  The second important parameter set was those defining the
background sky model.  Crowded fields result in a background sky level which
varies significantly across the field, and \dph\ can model this variation
using either a uniform gradient model or a modified Hubble profile (for the
latter see e.g.\ Binney \& Tremaine 1987).  We adopted the Hubble model since
it increased the detection rate by a factor of $\sim$3 due to the more
realistic background sky description.

These two sets of parameters were then hard-wired into the fine-tuned template
input list. Yet, there are two values that need to be specified separately for
each frame, namely estimates of seeing and background sky.  We created the
pipeline in such a way that it measures these two values directly from the
image and adds them to the parameter template to create the final input files.

\subsection{ Calibrating \dph's output to the AB and J2000 systems  }
\label{dph_analysis}

During its photometry routine, \dph\ outputs a list of detected objects with
several quantities, including the object positions and classification, total
and `aperture' magnitudes (with corresponding errors).  \dph's magnitudes are
given in the form of $-2.5\log{\mathrm{(DN)}}$, where DN are counts in units
of digital numbers.  To convert them to the AB\footnote{ SDSS photometry is
  intended to be on the AB system \citep{okegunn83,fukugita96}, by which a
  magnitude $0$ object should have the same counts as a source of $F_\nu =
  3631$~Jy. However, the photometric zeropoints are slightly off the AB
  standard. Nonetheless, according to the present estimate the (\g,\r,{\em i})
  band zeropoints are close to the AB system ($\sim 0.01$~mag). For details
  see www.sdss.org.  } magnitude scale (for details about the AB photometric
system see \citealt{fukugita96} and references therein) we used the parameters
from the SDSS tsField files to calibrate \dph\ counts:
\begin{equation}
\label{pog_mag}
m = -2.5 \log{ \left( \frac{ \mathrm{DN} }{t_\mathrm{exp} }\right) } - c_0 - 
         c_\mathrm{ext}\cdot m_\mathrm{air}
\end{equation}
where $t_\mathrm{exp}$ is the exposure time ($t_\mathrm{exp} =
  53.907456$~s), $|c_0|$ is the magnitude that yields
  $\mathrm{DN}/t_\mathrm{exp}$ of 1 at zero airmass, $c_\mathrm{ext}$ the
  extinction coefficient and $m_\mathrm{air}$ the airmass.  The values $c_0$,
  $c_\mathrm{ext}$, $m_\mathrm{air}$ with their uncertainties are extracted
  from the SDSS tsField file\footnote{The SDSS tsField files report
    $c_0<0$.}. 

A final magnitude correction is calculated via a set of `perfect' stars, those
   objects which best represent the PSF shape.  [To find these stars \dph\
   iterates through the given image starting at the highest level. At a 
   given level, it creates a typical stellar shape using all detected stars
   within this iteration. Then it fits this shape to a given object and
   determines the magnitude. This procedure is repeated at each level. If the
   shape {\em and} magnitude fits converge for a given object, it is 
   classified as a `perfect' star.]  The final correction is then calculated as
   the median offset between the total and 
aperture magnitude systems for these perfect stars.  The correction is applied
   to the complete object list, providing a final set of total AB magnitudes
   (calculated from the total fluxes derived from fitting a model PSF).

The AB magnitude errors are calculated
taking into account the errors of the variables used in 
eq.~[\ref{pog_mag}]. Those are then added to the median uncertainty 
in quadrature to calculate the final uncertainties. Thus, for any given object 
reported by \dph, the magnitudes are scaled to the AB system,
aperture corrected and the uncertainties
are reported. This is completed separately for each frame.
From the positions in pixel coordinates as given by \dph\ the J2000 right
ascension and declination for each object are recovered. 
The objects extracted in each frame are then matched between all bands
allowing for a maximum separation of $\sim2.8''$.
Thus, a final
stellar list with photometric magnitudes and uncertainties in \g, \r\ and {\em
  i} is obtained. In order to compare this photometry with the existing SDSS
data, the match algorithm is used to find common stars between the two
datasets.   

\section{Testing \dph}
\label{sec:tests}

In this section we describe tests performed on SDSS images using our automatic
package (see previous section). Our aim was to constrain the photometric 
accuracy and the detection efficiency of the pipeline, in particular for
crowded fields where the SDSS photometric software, \ph, gives poor
results (see Fig.~\ref{fig:leosdss}). The results below demonstrate that our
DoPHOT-based pipeline can extract high quality photometry from SDSS  
crowded-field images.

\subsection{ Photometry }

\begin{figure}[h!] 
\centering
\includegraphics[bb=20 70 592 679, width=\columnwidth]{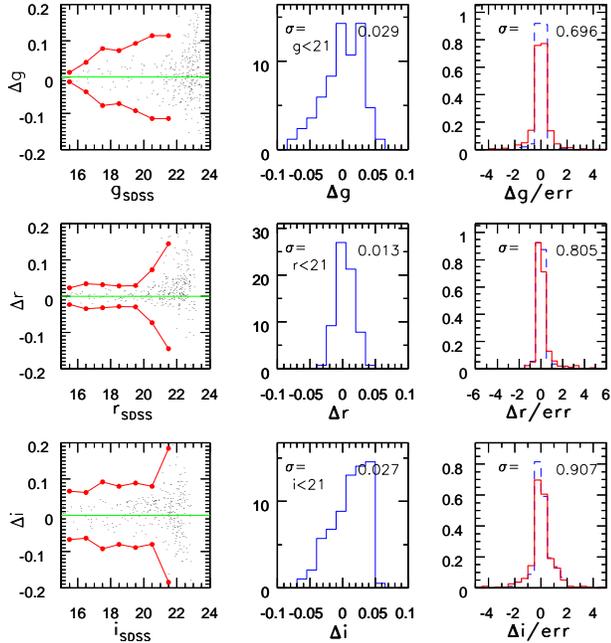}
\caption{The comparison of \dph\ and SDSS magnitudes in an un-crowded field
  (run 3631, camcol 3, rerun 40, field 236) in the surrounding of \leo. 
The first column displays the magnitude difference as a function of magnitude;
the large symbols connected by lines show the $\pm 3 \sigma$ envelope. The
distribution of magnitude differences at the bright end is displayed in the
middle column. The last column shows the distribution of the magnitude
differences normalized by the expected errors of \dph's and \ph's photometry
added in quadrature at the bright end (solid line) and for the full sample
(dashed line). }
\label{dph_sdss}
\end{figure}

The accuracy of \dph's (\g, \r, {\em i}) magnitudes relative to SDSS imaging
magnitudes is summarized in Fig.~\ref{dph_sdss} for stars in an un-crowded
field. The left column shows the magnitude difference as a function of SDSS
magnitude. The middle column shows the histogram of the magnitude difference
at the bright end ($\mathrm{mag}<21$).  The distribution of the magnitude
difference is symmetric in the \r\ band, while the \g\ and {\em i}\ bands show
slight systematic trends caused by the differences between the two pipelines,
\ph\ and \dph.  One possible source for this skewed behavior may be the
  aperture parameters which were used as an input for \dph, and were
  optimized in the r band. It is also possible that the different sky
  background estimations used in the two pipelines contribute to these
  effects.  The right column in Fig.~\ref{dph_sdss} displays the magnitude
differences normalized by the expected errors determined by adding the SDSS
and \dph\ errors in quadrature.  Note, that the errors include both
  photon noise errors and other systematic uncertainties. The mean equivalent
  Gaussian width, $\sigma$, indicated in each panel, is close to unity and
  demonstrates that
various possible errors in extracting counts do not exceed the noise expected
from photon statistics.


\subsection{ Detection efficiency }
\label{sec:cmpl}

In order to estimate the accuracy of \dph's photometry and the efficiency
of recovering stars as a function of both magnitude and stellar density,
we performed artificial star tests in crowded and un-crowded fields.
The tests were performed separately for (\g, \r, {\em i})\ bands.
Synthetic stars were generated using the \dph\ PSF model, which consists of
similar ellipses of the form: 
\begin{eqnarray}
\label{dph_psf}
I(z) & = & I_0 \left[ \sum_{k=0}^{3} \frac{1}{k!}z^k \right]^{-1}\\
z & = & \frac{1}{2}\left(x^2 + y^2\right)  
\end{eqnarray}

\begin{figure}[h!]
\centering
\includegraphics[bb=14 14 303 321, width=\columnwidth]{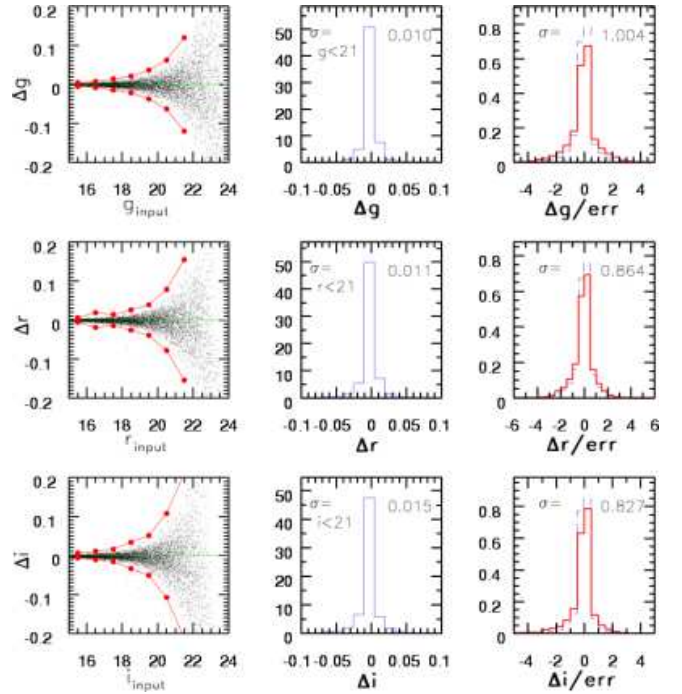}
\caption{ Artificial star test results for an un-crowded
field (same field as in Fig. \ref{dph_sdss}). The first column displays the 
magnitude difference of injected and extracted magnitudes as a function of 
magnitude; the line connected with symbols represents the $\pm 3 \sigma$
envelope. The middle column shows the distribution 
of magnitude differences for mag$<21$.
The last column displays the distribution of  magnitude differences normalized
by the expected errors at the bright end (solid line) and for the full sample
(dashed line). }
\label{dph_artf_nocrowd}
\end{figure}

\begin{figure}[h!]
\centering
\includegraphics[bb=14 14 303 321, width=\columnwidth]{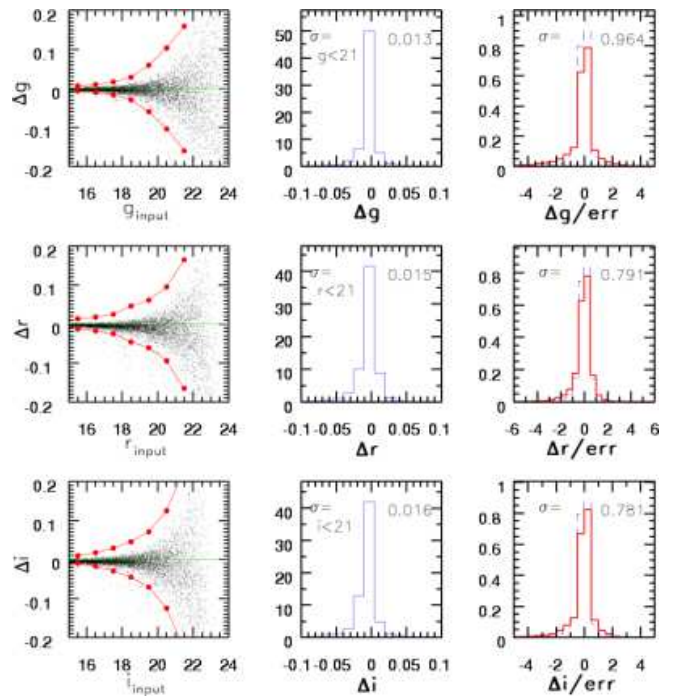}
\caption{ Same as Fig. \ref{dph_artf_crowd}, but for a crowded field 
encompassing a part of \leo\  (run 3631, camcol 3, rerun 40, field 238).}
\label{dph_artf_crowd}
\end{figure}

To avoid over-crowding (i.e.\ blending), the artificial star tests were done 
via Monte Carlo simulations by adding 100 randomly distributed synthetic
stars with an appropriate FWHM into each frame. 
To assess the photometric accuracy of the recovered stars, the procedure was 
repeated 100 times. The photometry was then extracted utilizing our pipeline as
described in Sec.~\ref{sec:datared}. 
In Fig.~\ref{dph_artf_nocrowd} we summarize the photometric accuracy for the
same un-crowded field shown in Fig.~\ref{dph_sdss}. The results for a crowded
field  are shown in Fig.~\ref{dph_artf_crowd}.  
The magnitudes are reproduced with the expected accuracy 
($\sigma \approx 0.014$), 
with no systematic errors in the overall magnitude scale (the width of the
distribution in the last column is very close to unity).
The slight systematic offset in the magnitude differences of
$\approx 5\x 10^{-4}$ and $\approx 5\x 10^{-3}$ for the un-crowded and
crowded field, respectively, is due to the aperture correction (the `aperture
box' parameters were set to fit SDSS magnitudes in un-crowded regions). Since
the offset is of order of $10^{-3}$~mag, while the magnitudes are accurate to
$10^{-2}$~mag, the discrepancy is negligible. We conclude that our
photometry is sufficiently accurate for robust studies of, for example, nearby
galaxies resolved by SDSS.


\begin{figure}[h!]
\centering
\includegraphics[bb=20 470 592 679, width=\columnwidth]{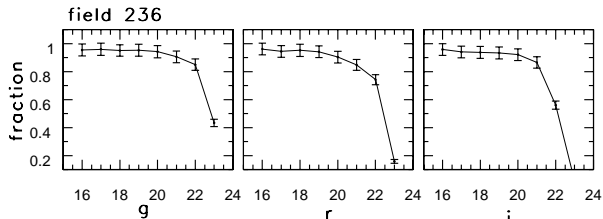}
\caption{ The fraction of recovered artificial stars (completeness) 
  as a function of magnitude for an un-crowded field surrounding \leo\ (see
  text for details). Note that the completeness starts declining rapidly for
  $\mathrm{mag}\gtrsim 21$.  } 
\label{compl} 
\end{figure}

For a thorough analysis of star counts, luminosity functions, surface density
profiles etc., one needs to understand the limits of the source extraction
software. It is already well known that detection efficiency declines with
stellar brightness. However, since this work is focused on crowded fields, it
is essential to understand how the number of retrieved stars depends on the
local stellar density.  To quantify these dependences we define the {\em
  completeness} as the ratio of the number of artificial stars extracted by
\dph\ to the number of artificial stars added to the frame,
$n_\mathrm{output}/n_\mathrm{input}$.  Fig.~\ref{compl} shows the completeness
as a function of magnitude for different SDSS fields.  We consider a star as
{\em recovered} if i) the position as extracted by \dph\ matches the position
of the inserted artifical star (within a box of $2''$ on a side centered on
the star) and ii) the classification for that object given by \dph\ is a
'perfect' star. The uncertainties in Fig.~\ref{compl} were derived from
Poisson errors for the input and output star counts in $1$~mag wide bins.
Therefore, they reflect the uncertainties due to the low number of artificial
stars injected rather than the intrinsic uncertainty of the completeness as a
function of magnitude.  The fraction of recovered stars falls below $90\%$ at
magnitudes fainter than $20-21$ (depending on the filter). For reference, the
quoted SDSS $95\%$ completeness is in the range of $21.3-22.2$~mag for the
(\gmag,\rmag,\imag) bands (using \ph); note, however, that this value is only
representative for un-crowded regions.

\begin{figure}[h!]
\centering
\includegraphics[bb=54 450 486 630, width=\columnwidth]{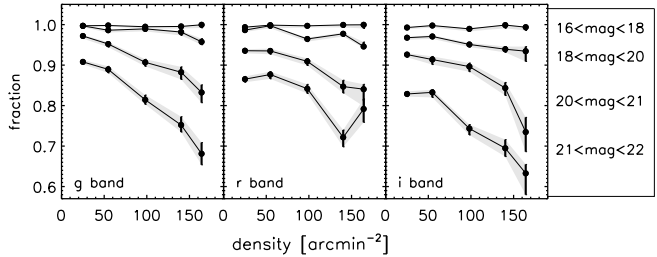}
\caption{ The fraction of recovered artificial stars (completeness) 
  as a function of stellar density for different magnitude bins. The panels
  present the combined artificial star tests ran on fields 237 and 238 (run
  3631, camcol 3, rerun 40). The band for which we show the derived
  completeness is indicated in each panel, and each curve represents the
  completeness for one magnitude bin, which is indicated in the legend on the
  righ-hand side. Note that the completeness is essentially not a function of
  stellar density till $\mathrm{mag}\sim20$. }
\label{compl237}
\end{figure}

Our next aim was to measure the completeness as a function of stellar density.
 The following tests were performed on two crowded fields at the center of
 the \leo\ dSph (run 3631, camcol 3, rerun 40, fields 237 and 238).  Each
 field was divided into cells of $50 \times 50$ square pixels in area, and the
 stellar density in each cell was then measured {\em prior} to artificial star
 injection.  The artificial stars were then placed in the field and recovered
 using the pipeline, and the completeness fraction was measured for each cell.
 To ensure we obtained accurate completeness measurements, the Monte Carlo
 simulations described above were executed for 1000 iterations.  Thus, $10^5$
 stars were randomly placed in each individual (\g, \r, {\em i})\ frame. The
 combined results are displayed in Fig.\ \ref{compl237}.  Note that
 we show the completeness for four magnitude bins. The crowding at the center
 of \leo\ has no significant effect on stellar recoverability for stars
 brighter then $\sim$20th magnitude.  Stellar crowding reduces the
 completeness at fainter magnitude bins, inflicting a $10-30\%$ loss for
 stellar densities up to $\sim$$200$ stars/arcmin$^{2}$.

 In Table \ref{tab:compl} we summarize the completeness as a function of both
 magnitude and stellar density for the {\em i}\ band.  In latter sections we
 will use these results to correct star-counts and flux integrals for
 incompleteness introduced by \dph.


\begin{deluxetable*}{c|ccc|ccc|ccc|ccc}[h!]
\tablecaption {{\em i} band completeness as a function of magnitude and stellar
   density 
   \label{tab:compl}}
\tablewidth{0pt}
\tabletypesize{\tiny}
\tablehead{
\colhead{density} & 
\colhead{fraction [\%]} & 
\colhead{ upper } &
\colhead{ lower } &
\colhead{fraction [\%]} &
\colhead{upper} &
\colhead{lower} &
\colhead{fraction [\%]} & 
\colhead{upper} &
\colhead{lower} &
\colhead{fraction [\%]} &
\colhead{upper} &
\colhead{lower} \\ 
\colhead{$[\mathrm{arcmin}^{-2}]$} & 
\colhead{$i\in(16,18)$}  & 
\colhead{ error } &
\colhead{ error } &
\colhead{$i\in(18,20)$}  & 
\colhead{ error } &
\colhead{ error } &
\colhead{$i\in(20,21)$}  & 
\colhead{ error } &
\colhead{ error } &
\colhead{$i\in(21,22)$} &      
\colhead{ error } &
\colhead{ error }   
}
\startdata
25.0 &
99.28 &
0.02 &
0.41 &

96.78 &
0.19 &
0.51 &

92.60 &
0.37 &
0.67 &

82.87 &
0.63 &
0.66 \\

54.7 &
99.77 &
0.01 &
0.48 &

97.08 &
0.27 &
0.59 &

91.38 &
0.76 &
1.24 &

83.27 &
0.64 &
1.26 \\

98.1 &
98.96 &
0.12 &
0.62 &

95.10 &
0.40 &
0.64 &

89.63 &
0.89 &
1.26 &

74.33 &
0.98 &
1.59 \\

140.2 &
99.89 &
0.02 &
0.78 &

93.92 &
0.66 &
0.78 &

84.34 &
1.39 &
1.85 &

69.47 &
2.13 &
2.06 \\

164.2 &
99.37 &
0.05 &
00.83 &

93.44 &
1.14 &
2.54 &

73.47 &
3.62 &
4.87 &

63.27 &
2.21 &
5.29 \\

\enddata \tablecomments{ The first column lists the mean density for which the
  completeness was measured. For the given stellar density range the remaining
  columns designate the fraction of recovered stars for different {\em i} band
  magnitude bins and the upper and lower errors (in \%). The uncertainties are
  Poisson errors of the median completeness for a given magnitude and density
  bin. The fractions presented here are plotted in Fig.~\ref{compl237}.  }
\end{deluxetable*}

\section{ Photometric properties of \leo\  }
\label{sec:leophotom}
In this section we present the construction of a high-contrast map of \leo\ 
utilizing the color-magnitude information, briefly describe the main features
of the color-magnitude diagram (CMD) of \leo, and compare them to previous
results.

\subsection { Photometric selection of \leo\  candidates }
\label{sec:mask}

\begin{figure}[h!]
\centering
\includegraphics[bb=14 14 399 263, width=\columnwidth]{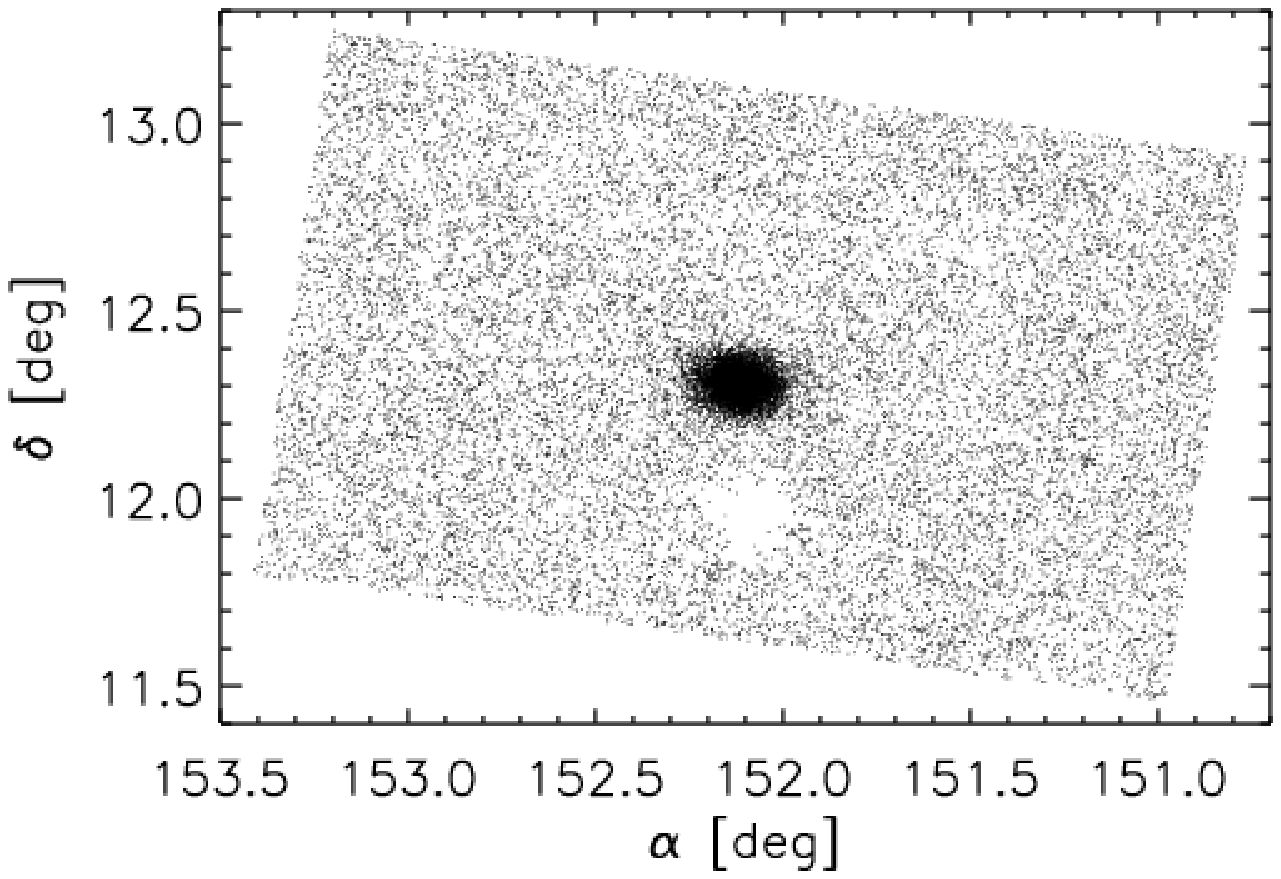}\\
\includegraphics[bb=14 14 399 263, width=\columnwidth]{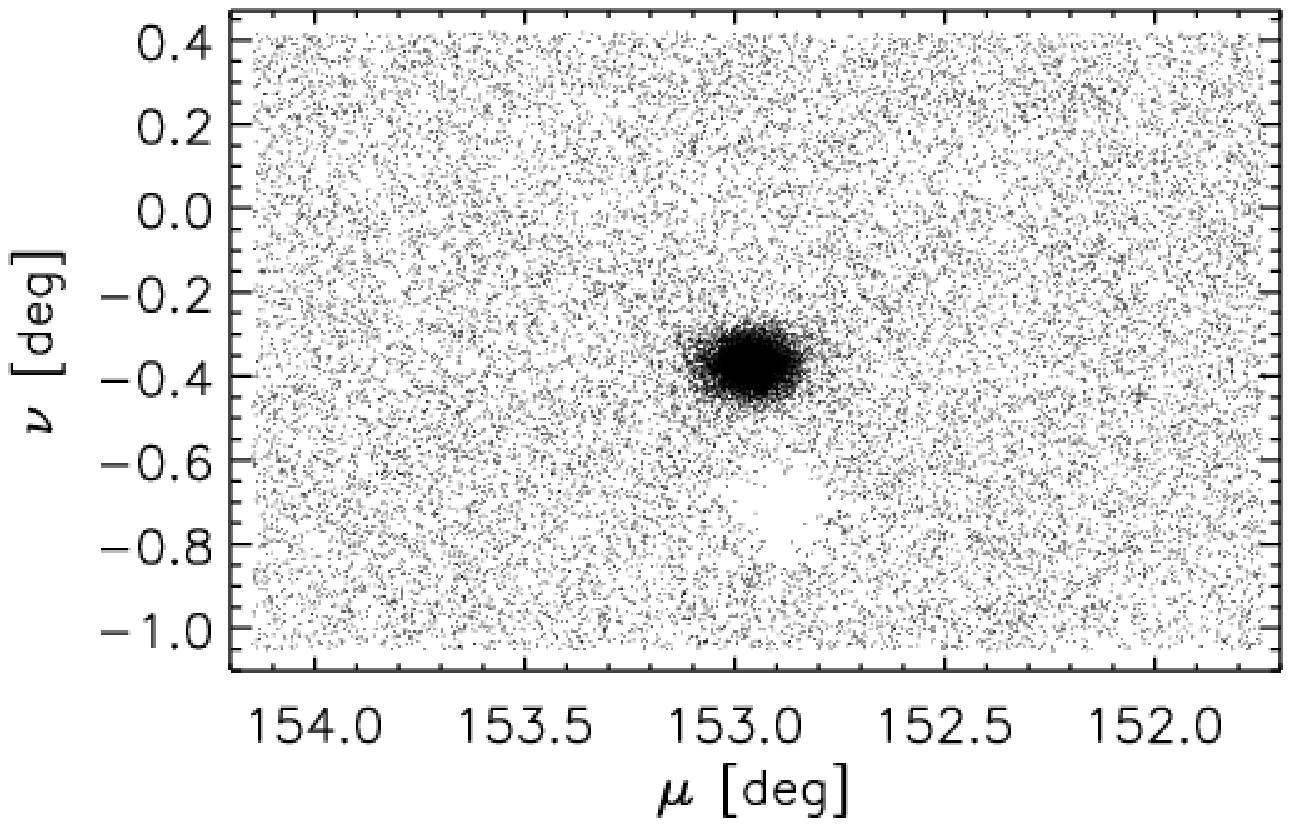}
\caption{ Spatial distribution of objects classified as 'perfect' stars by
  \dph. The shown area covers $\sim3.55\sqdegs$ around \leo. The \dSph\ is
  visible in the central region of the panel. The top panel shows the
  distribution in right ascension and declination (J2000), the bottom in great
  circle coordinates (used for scanning the sky by the SDSS). The empty
  area in both panels, south of \leo, is due to contamination by the first
  magnitude foreground star Regulus ($\alpha$ Leonis).  }
\label{fig:leo}
\end{figure}

Table \ref{tab:fields} summarizes the SDSS great circle scans which encompass
$\sim3.55\sqdegs$ around the \leo\  \dSph\ galaxy (see Fig.~\ref{fig:leo}). The
photometry extraction was performed as described in
\S \ref{sec:datared}.    An examination of the \citet{schlegel98}
dust map shows that extinction is not a significant problem for this 
object; reddening varies by approximately 0.01 mag across the \leo\ field.
The mean value at the centre of the dSph is $E(B-V) = 0.036$, and thus the
entire photometric dataset was corrected using this value.  The empty area
south of \leo\ in Fig.~\ref{fig:leo} is a result of contamination by the 1st
magnitude star Regulus. This and other regions around bright stars are
excluded in the further analysis.

\begin{deluxetable}{@{}c @{     }|c @{  } |c @{  }| c @{  } }
\tablecaption{Specifications for SDSS images utilized in this
  paper\label{tab:fields}}   
\tablehead{ \colhead{Run} & \colhead{Rerun} & 
         \colhead{Camcol} & \colhead{Fields} } 
\startdata
   3631 & 40 & 2 & 231-242 \\
   3631 & 40 & 3 & 231-242 \\
   3631 & 40 & 4 & 231-242 \\
   4338 & 40 & 1 & 54-69 \\
   4338 & 40 & 2 & 54-69 \\
   4338 & 40 & 3 & 54-69 \\
   4338 & 40 & 4 & 54-69 \\
\enddata
\tablecomments{Specifications for SDSS images that were used for the scope of
  this paper. They encompass an area of$\sim3.55\sqdegs$ around \leo. Each
  field is characterized with a set of four numbers (i.e. run,  
  rerun, camcol, field) which make it unique.}
\end{deluxetable} 

\begin{figure}[h!]
\centering
\includegraphics[bb=14 14 274 315, width=\columnwidth]{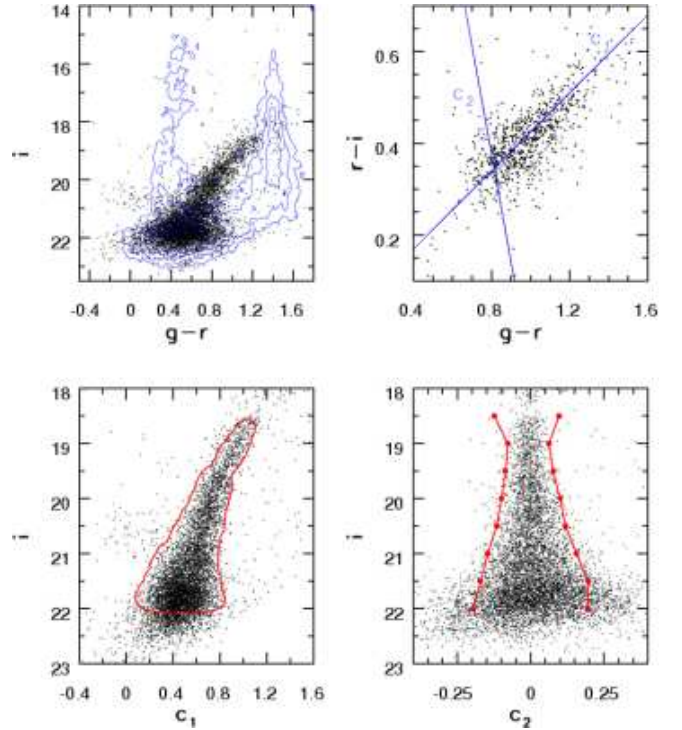}
\caption{ Color-magnitude and color-color diagrams of stars in the region
  around \leo. In the top left panel dots represent the central region of
  \leo, i.e. stars within an ellipse with a semi-major axis of $6'$ in the
  central region of \leo. Overlaid contours show the distribution of field
  stars obtained from stars beyond an ellipse with a semi-major axis of $30'$
  centered at \leo\ (see text for details). The magnitudes are corrected for
  reddening.  The top right panel shows the $r-i$ vs.\ $g-r$ diagram for
    stars with $i>21$, which were used to obtain the principal component axes
    ($c_1$, $c_2$), which are indicated in the panel (see
    eqs.~[\ref{eq:c1}]~\&~[\ref{eq:c2}] and text for details). Note that the
  position of the crossing point of the principal axes was shifted for
  clarity.  The bottom panels show the color-magnitude diagrams for the stars
  in the inner \leo\ region (dots) utilizing the principal component
    colors. The contour in the bottom left panel indicates the optimized
  photometric filter for the selection of potential \leo\ members. The
  connected large points in the bottom right panel indicate the $\pm2\sigma$
  envelope of the $c_2$ vs.\ $i$ distribution.  }
\label{fig:CMDs}
\end{figure}

In order to determine the photometric characteristics of the \leo\  population,
we selected stars within an ellipse of $6'$ semi-major axis (approximately
half the limiting radius) 
in the central region of \leo. The distribution of field stars was obtained
from stars beyond an ellipse with a semi-major axis of $30'$. 
The color-magnitude diagram (CMD) for stars in the central
region of \leo\ is shown in the top left panel in Fig.~\ref{fig:CMDs}, where
the field star distribution is overlaid in contours. As 
anticipated, the  distribution of field stars shows two main populations: the
bluer halo stars ($ g-r\approx 0.4  $) and the redder ($ g-r\approx 1.35  $)
thick disk stars.

To robustly study the \dSph\ galaxy, we needed to construct a map 
which maximized the number of \leo\ stars and minimized the level of field
star contamination.  In order to do this, we followed the method of
\citet{grillmair95}, which was refined for SDSS photometry by \citet{oden01}
in their analysis of the Draco dSph. Stars within the inner 
region 
of \leo\ form a tight correlation in the $g-r$ vs.\ $r-i$ plane. This is
illustrated in the top right panel in Fig.~\ref{fig:leo}. To quantify the
width of this distribution we defined a set of principal axes ($c_1$, $c_2$)
where $c_1$ measures the position along the $g-r$ vs. $r-i$ locus, and $c_2$
the position perpendicular to it:  

\begin{eqnarray}
\label{eq:c1}
    c_1 = 0.920\cdot (g-r) + 0.391\cdot(r-i) \\
    c_2 = -0.391\cdot(g-r) + 0.920\cdot(r-i)
\label{eq:c2}
\end{eqnarray}
The bottom panels in Fig.~\ref{fig:CMDs} show the CMDs for stars within the
inner \leo\ region using the principal colors. Since $c_2$ is essentially a
measure of the photometric dispersion, we imposed a cut of $c_2\leqslant
2\sigma_{c_2}$ to \leo\ stars. Here $\sigma_{c_2}$ designates the standard
deviation of the $c_2$ distribution measured as a function of $i$ magnitude
(see bottom right panel in Fig.~\ref{fig:CMDs}). We imposed an 
additional magnitude cut of $i < 22$ to the stars in the analyzed area
since the completeness decreases significantly at fainter magnitudes and the
photometric uncertainties are much higher.

Following the method of \citet{oden01}, we constructed CMD functions for the
field and \leo\ populations.  The $(c_1,i)$ CMD was divided into a series of
cells, where the cell dimensions were $0.09$ mag in color and $0.35$ mag in
$i$-magnitude.  Additionally, each cell was separated from its `neighbors' by
0.015 mag in color and 0.05 mag in $i$.  This level of overlap between
neighboring cells ensured the CMD function was smooth and continuous.  As
described above, the \leo\ stellar population was selected from the inner $6'$
of the dwarf, while the field population was taken from the area beyond a
radius of $30'$.  This second value is $\sim$$2.5$ times the \leo\ tidal
radius, and hence the field area is not expected to contain any of the dSph's
stars.  The CMD function was then constructed for both the field and \leo\
populations. 

As described by \citet{grillmair95} and \citet{oden01}, we then derived a
`signal' for each CMD cell by comparing the \leo\ and field populations.  For
a given cell at position $(i,j)$, the signal was calculated as: 
\begin{equation}
s(i,j) = \frac{n_\mathrm{c}(i,j) - 
         gn_\mathrm{f}(i,j)}{\sqrt{n_\mathrm{c}(i,j) + g^2n_\mathrm{f}(i,j)}},
\end{equation}
where $n_\mathrm{c}(i,j)$ and $n_\mathrm{f}(i,j)$ describe the \leo\ (core)
and field populations respectively.  The factor $g$ is a scaling factor
defined as the ratio of the core to field areas.  We optimized the population
contrast $s$ using a threshold value $s_0$ such that $s\geqslant s_0$. The
value of $s_0$ was derived 
using eq.~[2] in \citet{oden01} in such a way that the contrast was
optimized for an annulus between $30'$ and $60'$ from the center
of \leo. The optimal color-magnitude filter mask defined by $s_0$ is
outlined in the bottom left panel in Fig.~\ref{fig:CMDs}. This filter removed
$\sim80\%$ of the field stars and enhanced the central stellar density contrast by
a factor of $\gtrsim4$.

\subsection { Color-magnitude diagram  }
\label{sec:cmd}

The morphology of the CMD presented here (see left panels in
Fig.~\ref{fig:CMDs}) agrees very well with those obtained by previous studies
(e.g.\ \citealt{rm91, demers94, gallart99a, gallart99b, held00, held01}).
The most prominent feature is a well defined red giant branch.
The red clump is visible in the range $ 0.1 < g-r < 1 $ and
$ 21 < i < 22.5  $. 
The stars in the horizontally extended region at $i \approx
20.5 $ and $ -0.3 < g-r < 0.3 $ correspond to anomalous Cepheids reported
first by \citet{wc84} and explored further by e.g.\ \citet{lee93}.

\section { Size and structure of \leo }
\label{sec:structure}

\subsection{ The center  }
\label{sec:center}

In this section our aim is to quantify the size and structure of \leo. Hence
it is important to constrain the center of the stellar surface-density
distribution of the \dSph. In order to do so we spatially binned the
color-magnitude filtered \leo\ candidate stars and fitted a $2$-dimensional
Gaussian to this distribution. To account for any artifacts due to our
procedure and to robustly estimate the uncertainties, we derived the center,
position angle (PA) and ellipticity\footnote{Ellipticity is computed as
    $1-b/a$, where a and b are the major and minor semi-axes of the
    ellipse.}  of the dwarf using bootstrap-resampling.  The
mean center in great circle coordinates and the corresponding errors are
$\mu=(152.970\pm0.003)^\circ$ and $\nu=(-0.365\pm0.002)^\circ$. The best fit
position angle for great circle coordinates is $PA = (-1.7\pm5.7)^\circ$ and
the ellipticity is $0.3\pm0.1$.
In the remainder of the paper we will take these coordinates as the center of
\leo\ and use the above derived PA and ellipticity values to deduce the
surface-density profile of the \dSph.

The right ascension and inclination of the ascending node of the great circle
scans of the area analyzed here are $95\deg$ and $15\deg$,
respectively. Hence, the center in J2000 coordinates is at
$\alpha=152.122\deg\pm0.003\deg$ and $\delta=12.313\deg\pm0.003\deg$,
i.e. $10^h08^m29.4^s\pm 0.8^s~+12^d18'48'' \pm 9'' $, the 
PA is $(-9.2\pm5.7)\deg$ and the ellipticity stays the same. Our PA and
ellipticity agree within the uncertainties with the values given in e.g.\
IH95.
The center derived here is roughly consistent with the values from past studies
(\citealt{zwicky61, gallou71, dresscon76, falco99}; see NED for a summary). It
is worth noting, however, that the positions of the 
center of \leo\ vary in the literature by $\sim20'$ and $\sim30'$ in right
ascension and declination, respectively.

\subsection { The size  }
\label{sec:king}

\begin{figure}[h!]
\centering
\includegraphics[bb=54 400 486 752, width=\columnwidth]{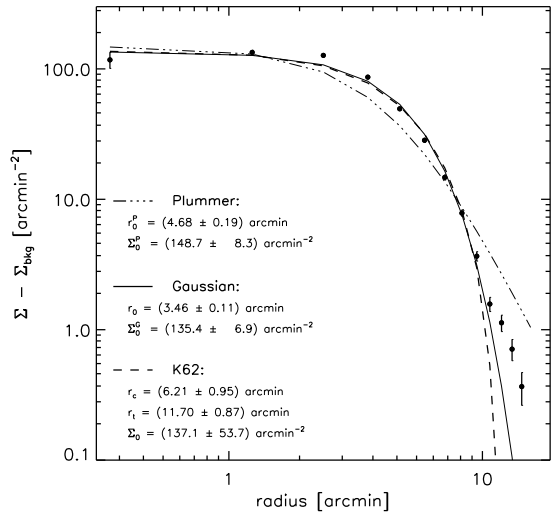}
\caption{ Radial profile of the completeness-corrected and
    background-subtracted stellar density for potential \leo\ members (dots).
    The error bars indicate quadratically combined errors of the background
    and stellar surface-densities, where both are taken to be dominated by
    Poisson and completeness uncertainties. Shown are the best-fit King (King
    1962; dashed line), Plummer (Plummer 1911; dash-dotted line), and
    Gaussian (solid line) profiles. We use the latter to estimate the mass of
    \leo\ (see text for details). The best fit parameters for all fitted
    models are indicated in the panel. Note that the Plummer profile gives
    the poorest fit to the data.  }
\label{fig:k62}
\end{figure}

In Fig.~\ref{fig:k62} we show the projected density profile of the
CMD-selected stars. The radial stellar surface-density was obtained by
computing stellar densities within elliptical annuli of $1.2'$ width starting
at the center of \leo\ (the center, ellipticity and position angle were chosen
as described above). Within each annulus we corrected for incompleteness (see
Tab.~\ref{tab:compl}), thereby accounting for the brighter photometric limit
of the crowded central region (see also Fig.~\ref{compl237}). The datapoints
in Fig.~\ref{fig:k62} display the completeness-corrected density ($\Sigma$)
above the background (also CMD selected and corrected for incompleteness;
$\Sigma_\mathrm{{bkg}} = (0.340 \pm 0.005)$~arcmin$^{-2}$) as a function of
the radius (i.e.\ the rms of the outer and inner semi-major axis for a given
annulus).  The projected background density was estimated using the
distribution of field stars beyond an ellipse of semi-major axis of $30'$ and
the same ellipticity as above centered at \leo. The error bars indicate
quadratically combined $\Sigma$ and $\Sigma_\mathrm{{bkg}}$ errors, where both
take into account completeness and Poisson uncertainties.

To quantify the size of \leo\  we fit an empirical King profile (King 1962; 
hereafter K62) to the stellar surface-density profile. The K62 profile is
given as: 

\begin{eqnarray}
\Sigma = k \left[ \frac{ 1 }{\sqrt{ 1 + (r/r_c)^2} } -  
             \frac{ 1 }{\sqrt{ 1 + (r_t/r_c)^2} }  \right]^2,\\
\Sigma_0 = k \left[ 1 - \frac{ 1 }{\sqrt{ 1 + (r_t/r_c)^2} }  \right]^2,
\end{eqnarray}

\noindent where $\Sigma$ is the stellar surface-density, $r$ the radius along
the major axis, $r_c$ and $r_t$ are the core and tidal radii, respectively,
and $\Sigma_0$ is the central surface-density. The best-fit K62 model, shown
in Fig.~\ref{fig:k62}, has a reduced $\chi^2$ value of $1.35$, and is
described with the following parameters: $\Sigma_0 = (137 \pm
54)$~arcmin$^{-2}$, $r_c=(6.21\pm0.95)'$ and $r_t=(11.70\pm0.87)'$.  Here the
uncertainties correspond to the standard deviation of the fitted parameters
based on the non-linear least square fit.  The core and tidal radii correspond
to $(460\pm75)$~pc and $(860\pm86)$~pc, respectively, assuming a distance of
$254^{+19}_{-16}$~kpc \citep{bellazzini04} to \leo. These parameters are
dependent on an accurate measurement of the \leo\ center.  It is possible that
the central coordinates derived in the previous section do not represent the
absolute minimum of the gravitational potential of \leo\ because the stellar
surface-density may not follow a $2$-dimensional Gaussian distribution. If we
repeat the derivation for determining the center of \leo\ described in the
previous section, but define the center to be the position where the stellar
surface-density has its maximum and fit the K62 model to those data, then the
core and tidal radii change by less than $2\%$.

Our best fit K62 parameters are somewhat different from the results
  reported in the literature (for an overview see IH95). While the tidal
  radius is in agreement with previous results (IH95 derive
  $r_t=(12.6\pm1.5)'$ and list previous results in their Table 8), our core
  radius is substantially higher.  $r_c$ is larger by $\sim50\%$ than the core
  radius reported in IH95.  This is most likely due to the photometry
  extraction package utilized here, which is specialized for crowded field
  photometry, as well as the applied robust completeness corrections, which
  are the largest in the central (i.e.\ most crowded) regions.

  We also find the best fit Plummer law \citep{plummer11}, $\Sigma =
  \Sigma_0^\mathrm{P} \left[ 1 + \left( r/r_\mathrm{P} \right)^2
  \right]^{-2}$, and show it in Fig.~\ref{fig:k62}. The reduced $\chi^2$ of
  the fit is $3.55$, and the central surface density and characteristic scale
  are $\Sigma_0^\mathrm{P} = (149 \pm 8)$~arcmin$^{-2}$ and $r_\mathrm{P} =
  (4.7 \pm 0.2)'$. The Plummer model is a less good fit to the data than the
  K62 profile.


In Fig.~\ref{fig:k62} we also show the best fit Gaussian profile to the data;
$\Sigma = \Sigma_0^{\mathrm{G}} \exp{\left[-r^2 / (2r_0^2)\right]}$. The
projected stellar density is well presented ($\chi^2=0.75$) by a Gaussian
out to a radius of $\sim12'$, with the best fit parameters of
$\Sigma_0^{\mathrm{G}}=(135\pm7)$~arcmin$^{-2}$ and $r_0=(3.5\pm0.1)'$. We
utilize this profile in Sec.~\ref{sec:ml} to model the mass of \leo.

\subsection{ Tidal tails?  }
\label{sec:tides}

It is clear from Fig.~\ref{fig:k62} that the stellar surface density falls off
less sharply than the K62 model. Such an excess of stars may indicate tidal
extension; on the other hand it may indicate that the (tidally-truncated) K62
model is simply a poor fit to the surface-density profile of a
tidally-undisrupted \dSph.  Some insight into this issue can be gleaned from
exploring the 2D density of stars in the outer parts of \leo.

\begin{figure}[h!]
\centering
\includegraphics[bb=54 400 486 680, width=\columnwidth]{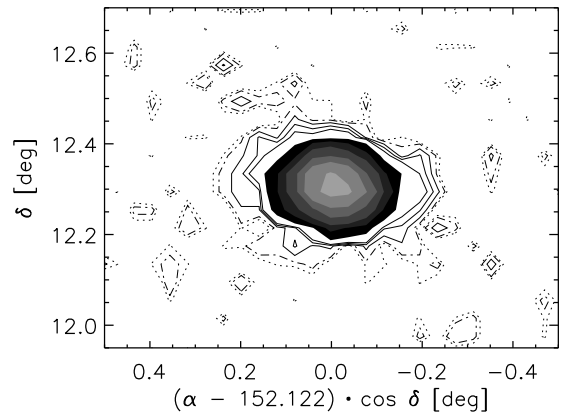}
\caption{ The spatial distribution of possible \leo\ candidates (see
  color-magnitude filter in the bottom left panel of Fig.~\ref{fig:CMDs}) with
  subtracted field contribution presented in contours. The contour levels
  correspond to surface densities derived in elliptical annuli of $1.2$' width
  (see text for details). The filled contours are equivalent to the surface
  densities well fit by the K62 profile (first 9 points in
  Fig.~\ref{fig:k62}). The solid and dotted contours correspond to densities
  poorly fit by the K62 model (the last 5 points in Fig.~\ref{fig:k62}; see
  text for details). The lowest density levels (dash-dotted and dotted
    lines) seem to show asymmetries.  The significance of these contours is
    however below $3\sigma$ in the background fluctuation of the field star
    population.  }
\label{fig:conts}
\end{figure}

We display the contour plot of \leo\ in Fig.\ \ref{fig:conts}.  This figure
 was constructed using the CMD-selected stars, where the background stellar
 density stated above in \S \ref{sec:king} has been subtracted.  Each contour
 level corresponds to a surface-density point in the radial profile.
 Thus, the apparent surface-density excesses at large radii seen in
 Fig.~\ref{fig:k62} can be directly related to the structures in Fig.\
 \ref{fig:conts}.  For clarity, the contour levels have been divided into two
 groups: those which are well fit by the K62 model (that is, the points in
 Fig.\ \ref{fig:conts} at $r \le 10'$; filled contours) and those which
 deviate significantly from the model ($r > 10'$; solid and dotted contours).
 The first four points beyond $r=10'$ are represented by solid contours, and
 they describe a roughly elliptical shape.  That is, although the radial
 profile of \leo\ appears slightly `inflated' in its outer regions,
 there is no evidence of a distorted structure (such as the `S'-shaped outer
 structure of the Ursa Minor dSph; \citealt{md01,palma03}).  This argues
 against any strong tidal disruption of \leo.

The two lowest contours (corresponding to the two outermost points in
 Fig.~\ref{fig:k62}) are displayed with dotted and dashed lines, respectively.
 A small surface density enhancement is visible to the north-east of
 \leo.  
 However, this possible excess of \leo\ stars is consistent with
 less than $3\sigma$ in the background fluctuation of the field star
 population.  Hence, within the limit of our data set, there is no clear
 evidence for a possible tidal disruption of \leo. This result is not
 surprising.  The relatively large luminous mass and Galactocentric distance
 of \leo\ argue against its displaying strong signs of tidal disruption.  Both
 \citet{byrd94} and \citet{peebles95} have modeled the orbit of \leo\ and
 argued that the galaxy had at most one encounter with a large Local Group
 galaxy in the past. This is consistent with its rather small tidal radius and
 no evidence for disruption beyond it. Independently, \citet{bowen97} showed
 that there is no evidence for tidally disrupted gas in \leo\, based on three
 QSO/AGN spectra which they utilized to search for absorption by gas within
 the halo of the \dSph.  

 We proceed in our analysis with the conclusion that \leo\ is not tidally
 disrupted, at least to a magnitude limit of $i=22$ and a stellar
 surface-density of $4\x10^{-3}$ of the central surface-density. Large-area
 coverage with deeper photometry would be needed to resolve whether there are
 tidal extensions of \leo\ at fainter levels.

\section{ Total luminosity, mass and mass-to-light ratio of \leo}
\label{sec:ml}

\subsection { Total luminosity }  

The CMD-filtered dataset was used to measure the luminosity function (LF) and
total luminosity of \leo.  First, we derived LFs for the field and core
regions of the dSph by counting the number of stars in bins of 0.15 magnitudes
down to the completeness limit of the survey, $i=22$.  In \S \ref{sec:center}
we derived the structural parameters of \leo, and these were used to define
the areas from which to draw the core and field populations.  The core
population was taken from within an ellipse with a semi-major axis of $12'$
(approximately the tidal radius), while the field population was defined to be
that beyond a semi-major axis of $30'$.  Both functions were corrected for
incompleteness using the estimates listed in Table \ref{tab:compl}.  The \leo\
LF was derived to be the difference between the core and field functions
(after scaling the field function to match the area from which the core
population was drawn). 

\begin{figure}[h!]
\centering
\includegraphics[bb=54 360 486 712, width=\columnwidth]{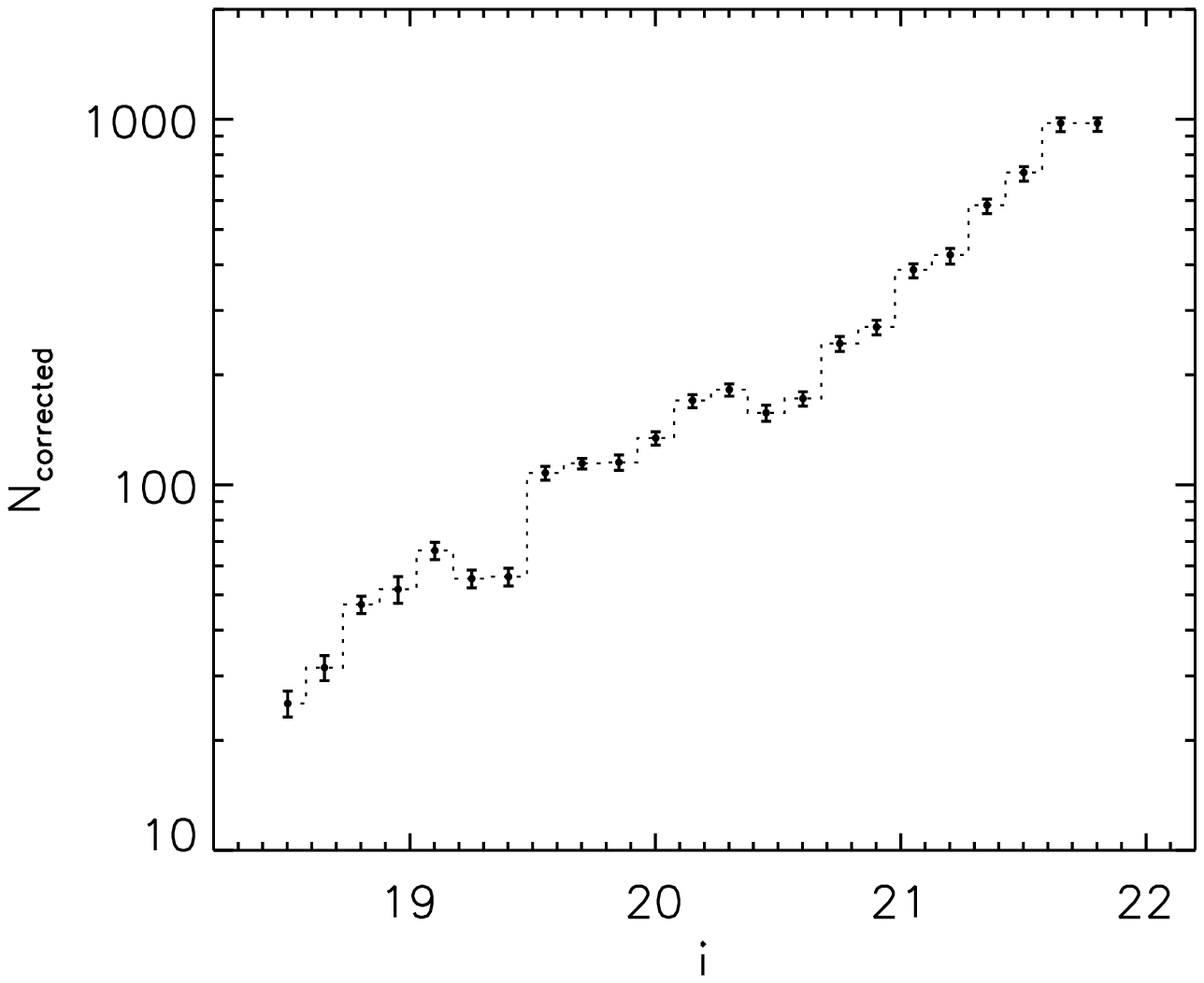}\\
\includegraphics[bb=54 400 486 722,width=\columnwidth]
                 {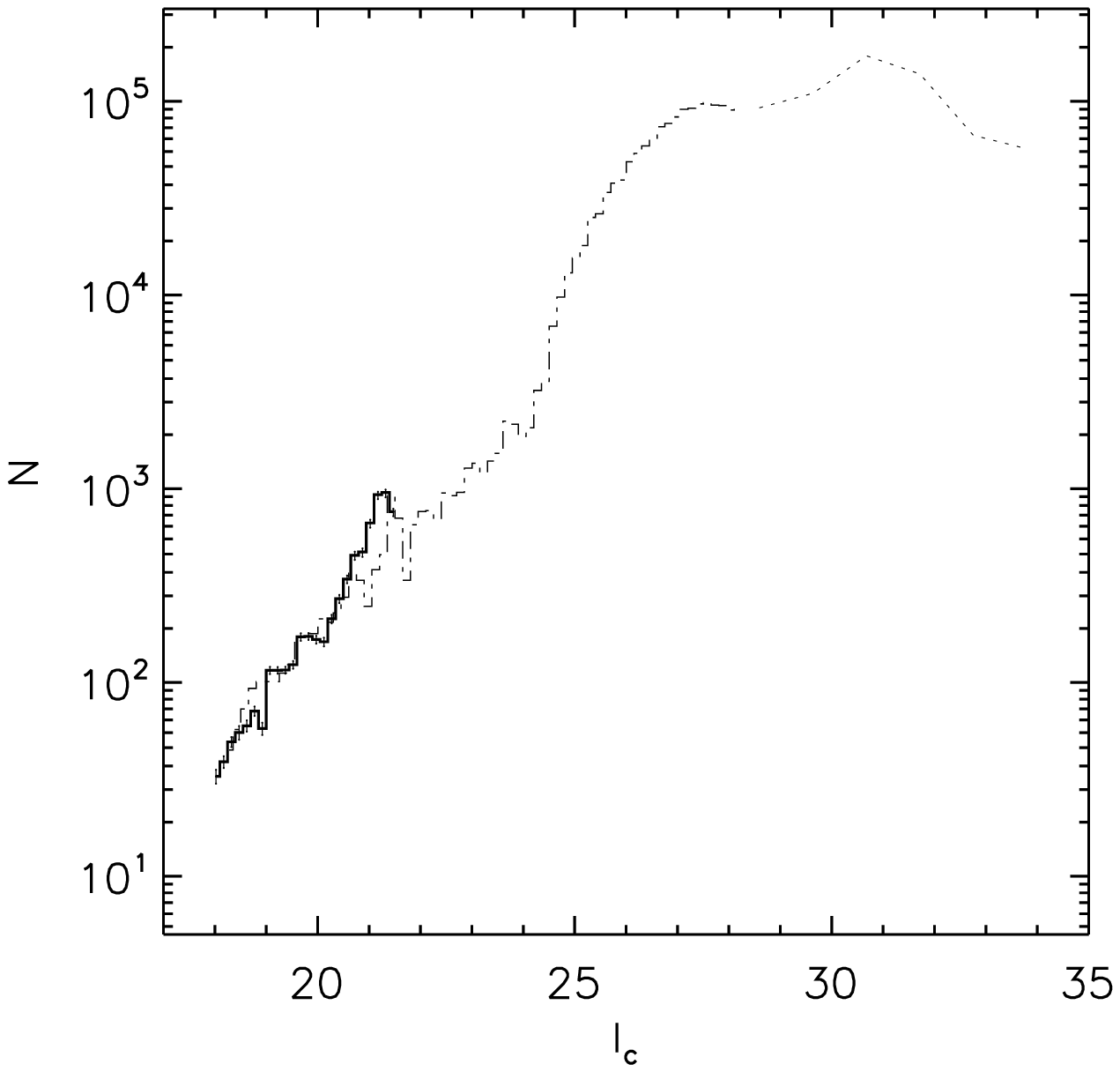}
\caption{ Top panel: \leo\ {\em i} band luminosity function (LF) corrected for
  the contribution of field stars and completeness (see text for details). The
  indicated errors correspond to Poisson and completeness uncertainties for
  field and \leo\ stars added in quadrature. The width of the magnitude bins
  is $0.15$~mag.  Bottom panel: Johnson-Morgan-Cousins \I\ band luminosity
  function of \leo\ (bold line) supplemented with the LF of the
  intermediate-metallicity globular cluster M5 (NGC5904; thin dash-dotted
  line) from \citet{sandquist96} and NGC6397 given by \citet{piotto97} (thin
  dotted line). The first LF was scaled to the distance and number counts (N)
  of \leo\ and the latter to the extrapolated LF of \leo\ (see text for
  details). }
\label{fig:LumFunct}
\end{figure}

Fig.~\ref{fig:LumFunct} (top panel) shows the completeness-corrected and
background-subtracted {\em i}\ band luminosity function of \leo. The error-bars
are the combination of Poisson uncertainties for the field and core
populations, including those taken from the completeness corrections. 
Our magnitude limit of $i=22$ excludes part of the red clump and the
horizontal branch
of \leo. \citet{held00} have shown that the morphology of the \leo\ 
horizontal branch is remarkably similar to that of the 
intermediate-metallicity globular cluster M5 (NGC5904;
\citealt{sandquist96}). Hence, to account for the missing flux of stars
fainter than our {\em i} band cutoff we supplemented the LF of \leo\ with the
one for M5 given in \citet{sandquist96}.  This step required a conversion of
the \leo\ luminosity function to the Johnson-Morgan-Cousin system via the
empirical  relations given in \citet{smith02}.  Distance moduli of
$22.02\pm0.13$ (\leo; \citealt{bellazzini04}) and $14.41\pm0.07$ (M5;
\citealt{sandquist96}) 
allowed us to place the LFs on the same magnitude scale.  Finally, the
functions were aligned by minimizing the difference in the overlapping 
magnitude range. To estimate the errors of such a procedure we scaled the two
LFs using the overlapping range (i) only to \I$=20.55$ (corresponding to
$i=21$) since our  mean photometric errors  beyond this limit are quite high,
i.e. $0.08\pm 0.02$, and; (ii) with no upper or lower limits.

Furthermore, we accounted for the missing flux beyond \I$=27.5$ using the
completeness corrected LF for NGC6397 given in \citet{piotto97}. We chose
NGC6397 since it has a metallicity comparable to the metallicity of \leo. We
used the distance modulus for NGC6397 given in \citet{piotto97} and
scaled it to the LF of M5 which has been previously scaled to the LF of \leo\
as described above. The resulting supplemented LF of \leo\ is shown in the
bottom panel of Fig.~\ref{fig:LumFunct}. It is in excellent agreement with the
LF of \leo\ presented in \citet{lee93} out to a magnitude of 
\I$\sim23$, beyond which their counts are dominated by incompleteness.  By
integrating over the luminosity function we calculated the total \I\
absolute magnitude of \leo\ to be $M_{I_c} = -12.03\pm0.14$.  Assuming a
solar absolute magnitude of $\MIsol = 4.14$ \citep{oden01} yields the
total luminosity of \leo\ of $L_{I_c} = (3.0\pm0.3)\x10^{6}\,\Lisol$. 

\subsection{ Total mass estimate and mass-to-light ratio }
\label{sec:model}

We estimate the central mass-to-light ratio, and on this basis the total
mass, of \leo\ using the Jeans equation for an isotropic spherical stellar
system (c.f. Binney \& Tremaine 1987):  

\begin{equation}
\frac{1}{\rho_*} \frac{ d\left(\sigma_*^2 \rho_*\right) }{ dr } = 
   - \frac{\mathrm{G} M\left(<r\right)} {r^2} 
\label{eq:jeans}
\end{equation}

\noindent where $r$ is the radius, $\sigma_*$ is the stellar velocity
dispersion, 
$\rho_*$ is the stellar density. The right hand term in eq.~[\ref{eq:jeans}]
represents the gravitational force of the system given in terms of the total
(i.e. stellar and dark matter) mass, $M$, within a radius $r$. 
For a simplified solution of eq.~[\ref{eq:jeans}] we assume that the
whole stellar system resides in a dark matter halo which dominates the 
mass. 

To solve eq.~[\ref{eq:jeans}] analytically for $\rho_{\mathrm{DMH}}$ we need to
find a simple model that describes the stellar volume-density distribution. In
\S \ref{sec:king} we have shown that the stellar surface-density
(that is, the projected stellar volume-density) profile is well fitted by
a Gaussian function (out to a radius of $\sim$$12'$). Since the projection of
a $3$-dimensional Gaussian distribution is again a Gaussian, we adopt
$\rho_*(r) = \rho_{*,0} \exp(-r^2/2r_0^2)$. We also make the ansatz that
$\sigma_*\approx\mathrm{constant}$, which leads to a solution 
of eq.~[\ref{eq:jeans}] with a central mass density:

\begin{equation}
\rho_{\mathrm{DM,0}} = 
                      \left(\frac{3}{4\pi}\right)
                      \frac{ \sigma_*^2}{\mathrm{G} r_0^2} 
\label{eq:mtotcent}
\end{equation}

\noindent \citet{mateo98} reported a velocity dispersion for the core of \leo\
of $(8.8\pm1.3)$~\kms\ based on 33 red giants in the \dSph. The average
angular separation of the observed stars from the center of the galaxy was
$1.9'$ with a largest separation of $3.5'$.  They have also shown that there
is no significant variation of the velocity dispersion with stellar type or
radius. Therefore, the ansatz that $\sigma_*$ is constant is a reasonable
assumption, at least in the central part of \leo. Hence, within a radius of
$\sim$$3.5'$ around the center of \leo\ the dark matter mass density 
is $\rho_{\mathrm{DM,0}}=(0.07\pm0.02)\,\Mo\,\mathrm{pc}^{-3}$ 
which corresponds to a central M/L
ratio of $\sim3$ in solar \I\ band units. Here the reported uncertainties are
formal statistical uncertainties. 

The inferred central mass density is comparable to the central density of
the Fornax \dSph\ \citep{mateo91} and Leo~II \citep{vogt95}. This is
consistent with the DM central densities for low-luminosity, gas-rich dwarf
irregulars (for a review see Mateo \etal\ 1991). On the other hand, it
is lower than the central densities inferred for other Local Group \dSph\
galaxies (see e.g.\ \citealt{mateo91} and \citealt{m98} for a review). Such a
result may argue in favor of DM in Local Group \dSph s being governed by 
Galactocentric distance; we discuss this in more detail in \S
\ref{sec:discussion}.   

With the limited kinematic data-set in hand we can only bracket the possible
M/L ratio of \leo\ (within the tidal radius) by investigating two limiting
cases: 

\noindent (i) We assume that the total mass profile follows the light
distribution. Although this assumption is not very likely
(e.g.\ \citealt{mateo91, kleyna01}), it provides a good constraint to the
minimum mass of \leo. The total mass is then given by: 

\begin{equation}
M_{tot} = 4\pi \int_0^\infty\rho_{\mathrm{mass}}(r)r^2dr = 
           \left(2\pi\right)^{3/2}\rho_{\mathrm{DM,0}}\,r_0^3
\label{eq:mfl}
\end{equation}

\noindent The total mass of \leo\ is then $ (1.7\pm0.2)\x10^7\,\Mo$, with
           quoted 
formal errors. Using the total luminosity of \leo\ derived in the previous
section we infer a mean \I\ band mass-to-light ratio of $\sim$$6$ in solar
units. 

\noindent (ii) We assume that the mass-density stays constant towards large
radii, i.e. $\rho_{\mathrm{mass}}(r)=\rho_{\mathrm{DM,0}}$. The total mass  
within a given radius is then:
 
\begin{equation}
M \left(r\right) = 4\pi
                   \int_0^r\rho_{\mathrm{DM,0}}\,r'^2dr' = 
                   \frac{4\pi}{3} \rho_{\mathrm{DM,0}}\,r^3
\label{eq:mcte}
\end{equation}

\noindent In this case the total mass of \leo\ within $r=12'$ is $(20 \pm
                   6)\x10^7\Mo$ 
and the mean mass-to-light ratio in the \I\ band is $\sim$$75$ in solar units.

Hence, for the two limiting descriptions of the distribution of mass in \leo\,
we infer an \I\ band M/L ratio of $6$ and $75$ in solar units,
respectively. If we adopt the mean $V-I_c$ color for \leo\ of roughly 0.6 (see
CMDs in \citealt{lee93} and \citealt{caputo99}), then the \V\ band M/L ratio
is about $5.5$ and $65$ in solar units, respectively, for the two limiting
cases. 

\section{ Discussion }
\label{sec:discussion}

We have estimated the mass-to-light ratio of \leo\ in the \I\ band to be in
the range of $6-75$ in solar units. The lower limit was calculated under the
assumption that mass follows light.  The upper limit represents the M/L ratio
within $\sim 12'$ of a system with a large dark matter halo extending far
beyond the limiting radius of the visible matter, where the dark matter halo
density is constant within the luminous limiting radius.  However, given the
current evidence concerning dark matter halos in dwarf galaxies, it is
possible to put constraints on which of these scenarios is the most likely.

If the distribution of mass follows the distribution of the visible component,
then the velocity dispersion of \leo\ would be expected to fall to zero at the
tidal radius. Nonetheless, past studies have shown that such a behaviour is
not typical for Local Group dSphs.  For example, \citet{kleyna01} have shown
that the radial variation of the velocity dispersion of the Draco \dSph\ is
flat within the uncertainties. Additional evidence for a flat velocity
dispersion profile is given by \citet{mateo91} for the Fornax \dSph.  The
evidence is similar for \leo.  Koch \etal\ (2007)
report that the velocity dispersion profile of \leo\ appears constant at
increasing radii.  This suggests that mass does not follow light in \leo\ and
favours option (ii).  We infer that the \leo\ M/L ratio is $\gg5$ in the
\V\ band. The inferred upper value of the M/L ratio in the \V\ band is
$\sim65$, yet this represents the M/L ratio only within $\sim 12'$; hence the
possibility that the true M/L ratio of \leo\ is even greater than this is not
ruled out.

It has been argued (for example, \citealt{klessen98}) that distortion due to
 the Galactic tidal field can reproduce the appearance of a system dominated
 by dark matter.  Under this scenario, the gravitational potential of the
 Galaxy has heated the internal structure of an orbiting satellite, thereby
 leading to an inflated velocity dispersion.  If we assume a {\em stellar}
 $M/L$ of 3 for \leo, then the velocity dispersion must be inflated by a
 factor of $4-5$ to produce an {\em apparent} $M/L$ of 65.  Given that \leo\
 is the farthest known dSph from the Galactic center, it would be surprising
 to find tidal forces strong enough to alter the internal kinematics
 to such a degree.  Indeed, our analysis found no evidence of tidal
 disruption.  In \S \ref{sec:tides} we demonstrated that the structure of
 \leo\ follows a smooth, elliptical shape to the limiting radius.  Thus, it
 seems unlikely that tidal forces are responsible for the large measured $M/L$
 value, and this result suggests that \leo\ is strongly dominated by dark
 matter. 

\section{ Summary }
\label{sec:summary}

We have presented an automatic photometric pipeline especially designed for
SDSS crowded-field images. The software performs extremely well on crowded
SDSS images, yielding high-quality photometry
with completeness above 80\% down to a magnitude of $\lesssim 21$ and stellar
density of up to $\sim200$~arcmin$^{-2}$. We have
extensively tested the detection efficiency and photometric accuracy of this
software   
(as a function of both magnitude and stellar density).  The pipeline was
applied to a region of $\sim 3.55\sqdegs$ around the dwarf spheroidal galaxy
\leo\ in the three bands \g, \r\ and {\em i}. 

We constructed a filter in colour-magnitude space targetting the \leo\ red
giant branch.  It removed $\sim80 \%$ of the foreground contamination, and
enhanced the central stellar density contrast by a factor of $\gtrsim4$.  We
find that the projected spatial structure of \leo\ is ellipsoid-like with core
and tidal radii (following an empirical King model) of $(6.21\pm0.95)'$ and
$(11.70\pm0.87)'$, respectively. This corresponds to $(460\pm75)$~pc and
$(860\pm86)$~pc, respectively.  The radial profile deviates slightly from the
King profile towards the outer regions, however there is no evidence for
extra-tidal structures (such as tidal tails) down to a magnitude of $i=22$ and
at a stellar surface-density of $4\x10^{-3}$ of the \dSph's central density.

The luminosity of \leo\ was measured by integrating the observed
completeness-corrected and background-subtracted flux of stars. 
We accounted for the missing flux to the faintest levels (\I=34)
supplementing the luminosity function with the LFs of the globular clusters M5
and NGC6397. The \I\ band absolute magnitude of \leo\ was measured to be
$M_{I_c} = -12.03\pm0.14$ and the total luminosity
$(3.0\pm0.3)\x10^{6}\,\Lisol$. 

We modelled the mass of \leo\ using the spherical Jeans equation and assumed
that the stellar velocity dispersion is isotropic and spatially constant. The
inferred central mass density of \leo\ is then
$(0.07\pm0.02)\,\Mo\,\mathrm{pc}^{-3}$.   
Assuming that mass follows light the total mass of \leo\ is
$(1.7\pm0.2)\x10^7\,\Mo$. This value is comparable to the total mass estimates
using the standard 'core' fitting method. Under the assumption that
the dark matter halo is extended and its density is spatially constant,
the total mass of \leo\ within the tidal radius of the visible component
(i.e. $12'$) is $(2\pm0.6)\x10^8\, \Mo$. Hence, the total mass to light ratio
for \leo\ in the \I\ band and in solar units is in the range of about 6 to
75, where the first and latter values correspond to the above quoted masses, 
respectively. In Sec.~\ref{sec:discussion} we argued that the mass-follows-light
assumption is not reasonable for the \leo\ system and therefore concluded that
the M/L ratio must be $\gg6$ in \I\ band solar units and possibly $>75$ if
a constant density DM halo would dominate the mass and extend further beyond
$12'$. 

\acknowledgments

The authors would like to thank the referee M.~Irwin for constructive
suggestions which helped to improve the paper.  VS thanks MM, DR and BB for
sincere and enormous support during the creation of this paper; Frank van den
Bosch and Robert Lupton for insightful discussions; and Andreas Koch and
collaborators for sharing their results with us prior to publication.

    Funding for the Sloan Digital Sky Survey (SDSS) has been provided by the
    Alfred P. Sloan Foundation, the Participating Institutions, the National
    Aeronautics and Space Administration, the National Science Foundation, the
    U.S. Department of Energy, the Japanese Monbukagakusho, and the Max Planck
    Society. The SDSS Web site is http://www.sdss.org/. 

    The SDSS is managed by the Astrophysical Research Consortium (ARC) for the
    Participating Institutions. The Participating Institutions are The
    University of Chicago, Fermilab, the Institute for Advanced Study, the
    Japan Participation Group, The Johns Hopkins University, the Korean
    Scientist Group, Los Alamos National Laboratory, the Max-Planck-Institute
    for Astronomy (MPIA), the Max-Planck-Institute for Astrophysics (MPA), New
    Mexico State University, University of Pittsburgh, University of
    Portsmouth, Princeton University, the United States Naval Observatory, and
    the University of Washington. 


\clearpage


\end{document}